\documentclass[prl,superscriptaddress]{revtex4-1}
\usepackage{amsmath,amssymb,latexsym,epsfig,graphics,epsf}
\usepackage{graphics}
\usepackage{float}
\usepackage{color} 
\newcommand{\eq}[1]{Eq.~(\ref{#1})}
\newcommand{\Eq}[1]{Equation~(\ref{#1})}
\newcommand{\fig}[1]{Fig.~\ref{#1}}
\newcommand{\Fig}[1]{Figure~\ref{#1}}
\newcommand{\be}{\begin{equation}}
\newcommand{\ee}{\end{equation}}

\begin{document} 

\title{Phase diagram of Kob-Andersen type binary Lennard-Jones mixtures}
\author{Ulf R. Pedersen}\email{ulf@urp.dk}
\affiliation{{\it Glass and Time}, IMFUFA, Department of Science and Environment, Roskilde University, P. O. Box 260, DK-4000 Roskilde, Denmark} 
\author{Thomas B. Schr{\o}der, and Jeppe C. Dyre}
\affiliation{{\it Glass and Time}, IMFUFA, Department of Science and Environment, Roskilde University, P. O. Box 260, DK-4000 Roskilde, Denmark} 
\date{\today} 

\begin{abstract} 
The binary Kob-Andersen (KA) Lennard-Jones mixture is the standard model for computational studies of viscous liquids and the glass transition. For very long simulations the viscous KA system crystallizes, however, by phase separating into a pure A particle phase forming an FCC crystal. We present the thermodynamic phase diagram for KA-type mixtures consisting of up to 50\% small (B) particles showing, in particular, that the melting temperature of the standard KA system at liquid density $1.2$ is $1.028(3)$ in A particle Lennard-Jones units. At large B particle concentrations the system crystallizes into the CsCl crystal structure. The eutectic corresponding to the FCC and CsCl structures is cut-off in a narrow interval of B particle concentrations around 26\% at which the bipyramidal orthorhombic ${\rm PuBr_3}$ structure is the thermodynamically stable phase. The melting temperature's variation with B particle concentration at two other pressures, as well as at the constant density $1.2$, is estimated from the simulations at pressure $10.19$ using isomorph theory. Our data demonstrate approximate identity between the melting temperature and the onset temperature below which viscous dynamics appears. Finally, the nature of the solid-liquid interface is briefly discussed. 
\end{abstract}  

\maketitle

Standard models are invaluable in physics by providing well-understood reference systems for testing new ideas. Well-known examples are the ideal gas model, the Ising model for critical phenomena \cite{goldenfeld,maniss}, and the $\phi^4$ scalar field theory for renormalization \cite{goldenfeld,amit}. In computational studies of viscous liquids and the glass transition \cite{bar00,fra02,ash03,fle05,IV,ped10} the Kob-Andersen (KA) binary Lennard-Jones (LJ) mixture has been the standard model for 20 years \cite{ka1}. The KA model is a mixture of 80\% large A particles and 20\% small B particles. The model is characterized by a strong AB attraction, which disfavors phase separation into a pure A phase by making it energetically costly. When the model crystallizes in very long computer runs, this nevertheless happens by phase separation into a pure A phase forming a face-centered cubic (FCC) crystalline structure \cite{hed09,tox09}. This may be contrasted to the Wahnstrom 50/50 binary LJ mixture \cite{wah91} that has a much more complex crystal structure \cite{ped07,ped10b}.

In the KA mixture all particles have the same mass and interact via LJ pair potentials  $v(r)=4\varepsilon\left[(r/\sigma)^{-12}-(r/\sigma)^{-6}\right]$ truncated and shifted to zero at 2.5$\sigma$ with $\sigma_{BB}/\sigma_{AA}=0.88$, $\sigma_{AB}/\sigma_{AA}=0.8$, $\varepsilon_{BB}/\varepsilon_{AA}=0.5$, and $\varepsilon_{AB}/\varepsilon_{AA}=1.5$. The large $\varepsilon_{AB}$ favors stability towards phase separation. The parameters of the KA model were chosen to mimic the nickel-phosphor mixture \cite{ka1} that has a eutectic at 20\% P atoms \cite{sch09a}; eutectic mixtures \cite{zha12} are generally regarded as optimal for glass formation \cite{tur69}.

Crystallization of binary LJ models has been simulated by several groups \cite{hit99,lam01,lam01a,kof93}. In 2001 Wales and coworkers showed that the KA system possesses low-lying crystalline minima \cite{mid01}, confirming the prevailing view that the KA viscous liquid is not in true thermodynamic equilibrium \cite{ast09}. Two years later Fernandez and Harrowell reported rapid crystal growth for the 50/50 KA mixture \cite{fer03} and argued that for general KA-type mixtures, the lowest-energy ordered state consists of coexisting phases of single-component FCC and equimolar CsCl crystal structures \cite{jun11}. In 2009 Kob and coworkers reported that the standard 80/20 KA mixture crystallizes in two dimensions whereas the 65/35 composition does not \cite{bru09}. Royall and coworkers in 2015 showed that locally favored structures impede crystallization if they  do not tile space \cite{cro15}. In 2016 Bhattacharyya and coworkers showed that the entropic penalty for demixing is a non-monotonic function of composition with a maximum at a composition close to that of the standard KA model \cite{nan16}.

\begin{figure}
	\begin{center} 
		\includegraphics[width=0.5\textwidth]{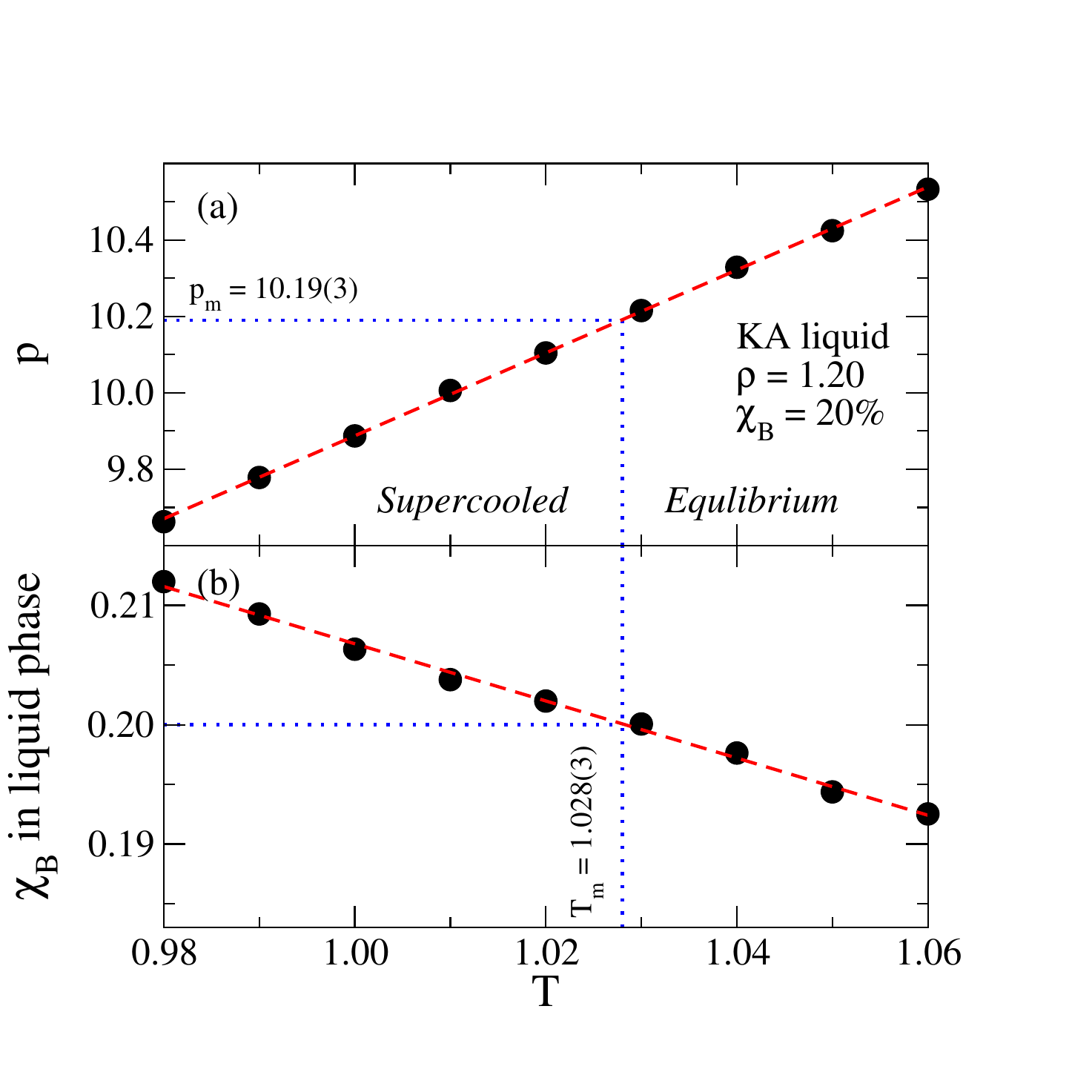} 
		\caption{\label{pLiquid} Determining the standard KA system's melting temperature and pressure when the liquid at coexistence has density 1.2.
		(a) The pressure of the KA liquid at density 1.2 as a function of temperature. (b) The fraction of B particles in the liquid phase of an equilibrated two-phase simulation at this pressure. The melting temperature of the density 1.2 KA liquid is seen to be given by $T_m=1.028(3)$ at pressure $p=10.19(3)$.}
	\end{center} 
\end{figure}

The below study utilized the software packages RUMD \cite{rumd15}, LAMMPS \cite{LAMMPS}, VMD \cite{VMD}, and a home-written code available at http://urp.dk/tools. The majority of the simulations involved 8000 particles, a few ones 10000 particles.

Most KA-model papers focus on density 1.2 in the LJ unit system used henceforth defined by the A particle parameters. We determined the melting temperature of the standard KA liquid at this density from two-phase simulations by proceeding as follows. First the pressure of the liquid was calculated as a function of temperature at density 1.2 (\fig{pLiquid}(a)). Next we determined the lattice parameters of pure A particle FCC crystals at the temperatures and pressures of \fig{pLiquid}(a) \cite{par81,mar94}. The final step was to simulate at each temperature and pressure a two-phase system composed initially of half KA liquid and half pure A crystal, choosing a box size compatible with the determined lattice constant and using a ``longitudinal'' barostat ensuring constant pressure perpendicular to the crystal-liquid interface. Over time the pure A crystal either grows or shrinks, which changes the ratio of A to B particles in the liquid phase. When equilibrium has been reached, the temperature is the melting temperature of that particular liquid composition \cite{bec06}. \Fig{pLiquid}(b) shows the fraction of B particles where the abscissa  is the melting temperature at the pressure identified in \fig{pLiquid}(a). For the standard 80/20 KA system we see that at pressure $p=10.19(3)$ and liquid-phase density 1.2, freezing occurs at temperature $1.028(3)$.

\begin{figure}
	\begin{center} 
		\includegraphics[width=0.44\textwidth]{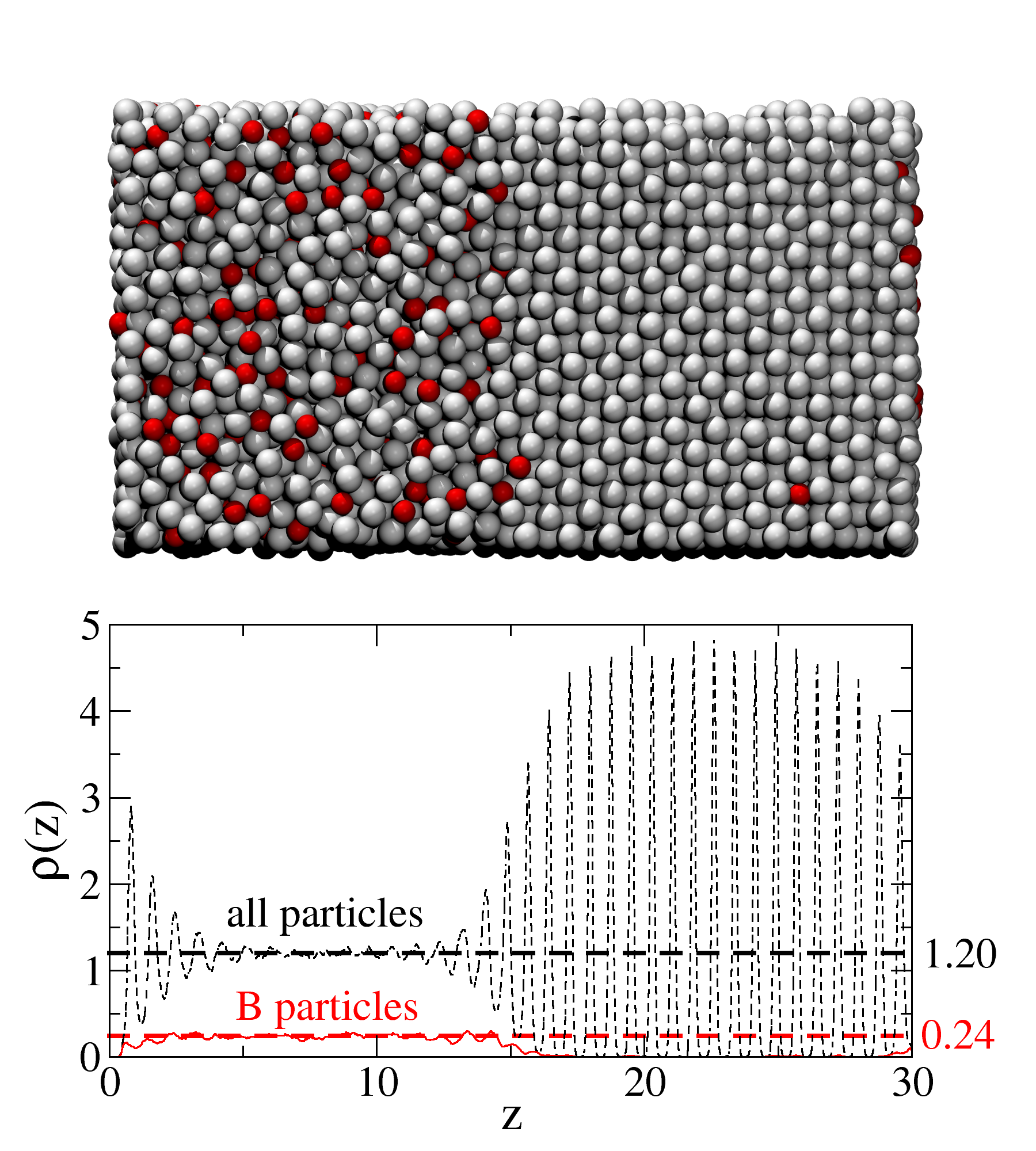} 
		\caption{\label{snapshot} Upper panel: Snapshot of a two-phase simulation in a periodic box at the coexistence temperature $1.028$ and pressure $p=10.19$ (A particle LJ units).  A particles are gray and B particles are red. The density of the liquid phase is 1.2. Lower panel: Overall density (black dotted line) and B particle density (red line) along the $z$ direction. The fraction of B particles in the liquid phase is $0.24/1.20=20$\%.} 
	\end{center}
\end{figure}

\Fig{snapshot} shows a snapshot of the two-phase equilibrium KA system at $T=1.028$ and $p=10.19$. A few B particles (red) have diffused into the pure A crystal, which is to be expected given the relatively high temperature. The lower panel shows the density of B particles (red) and of all particles (black), confirming that the liquid phase has density 1.2 and contains 20\% B particles.

\begin{figure}
	\begin{center}
		\includegraphics[width=0.5\textwidth]{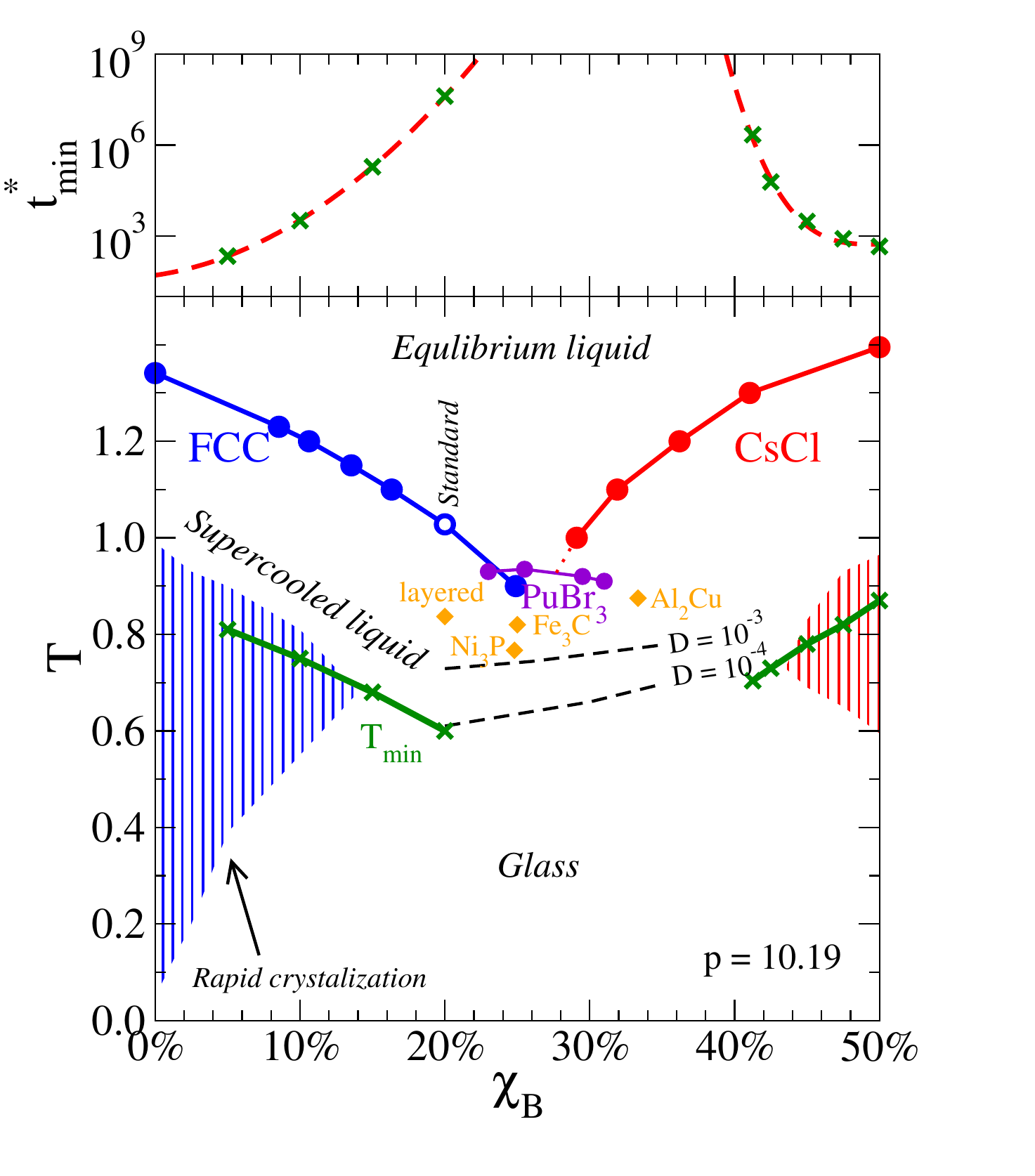} 
		\caption{\label{pd}Phase-diagram of KA-type mixtures at pressure $p=10.19$ where $\chi_B$ is the fraction of B particles in the equilibrium liquid phase. The solid blue line shows the melting temperature of the pure A particle FCC crystal structure, the solid red line shows the melting temperature of the CsCl structure, the short solid purple line shows the melting temperature of the PuBr$_3$ structure. Orange symbols give the melting temperature of other, less stable structures at selected compositions (see text). The two dashed black lines are isodiffusional lines. The green crosses show the temperature $T_{\rm min}$ of minimum crystallization time for a given $\chi_B$; the colored regions are where rapid crystallization occurs defined by $t^*<2\times10^5$ for 8000 particles. The upper panel shows the minimum crystallization time. Extrapolations based on third-order polynomials (red dashed lines) suggest that the best glass former is at a higher B particle composition than the eutectic.} 
	\end{center} 
\end{figure}

Having determined the pressure of the state point at which the KA liquid at freezing has density 1.2, we proceed to establish the composition-temperature phase diagram at this pressure (\fig{pd}). The phase diagram was arrived at by the above method in which the solid-liquid interface is ``pinned'' due to the composition constraint in the liquid. When the liquid has same composition as the crystal this constraint is absent, however, in which case we used the interface-pinning method \cite{ped13}. 

At small B particle concentrations the solid phase is a pure A phase FCC crystal. Upon increasing the B particle concentration, at some point the system crystallizes instead into the 50/50 CsCl structure of two interpenetrating cubic lattices of A and B particles \cite{fer03,nan16}. The eutectic of these two crystal phases is located around 26\% B particles, but interestingly there is a narrow region around the eutectic in which the bipyramidal orthorhombic ${\rm PuBr_3}$ structure is the thermodynamically stable phase. We also studied the Al$_2$Cu, Fe$_3$C, and Ni$_3$P structures shown by Fern\'{a}ndez and Harrowell to be of low energy at $T=0$ and $p=0$ \cite{fer04}, but found that their melting temperatures are all lower, compare \fig{pd} and Table \ref{tableTm}. In \fig{pd} ``layered'' denotes a structure of alternating layers of CsCl and pure A particle FCC crystals, known to have a low energy at $T=0$ \cite{mid01,fer03}. 

To investigate the dynamics of crystallization we utilized the following protocol for a range of compositions and temperatures ($p=10.19$): First, one equilibrates up to ten liquid configurations at the coexistence temperature. Subsequently, the temperature is changed to the desired value and one computes the crystallization time $t^*$, defined as the average time at which 10\% of the particles are in a crystalline environment (monitored by the crystal order parameter defined as the absolute value of the density's Fourier transform at the fundamental lattice k-vector \cite{lec08}). The blue and red areas in \fig{pd} are regions with rapid crystallization. As expected, the spontaneously formed crystals are FCC for 0\%-15\% B's and CsCl for 41\%-50\% B's. The green crosses show for each composition the temperature $T_{\rm min}$ at which the crystallization time is at a minimum, $t^*_\text{min}$. The upper panel shows $t^*_\text{min}$.

\begin{figure}
	\begin{center} 
		\includegraphics[width=0.5\textwidth]{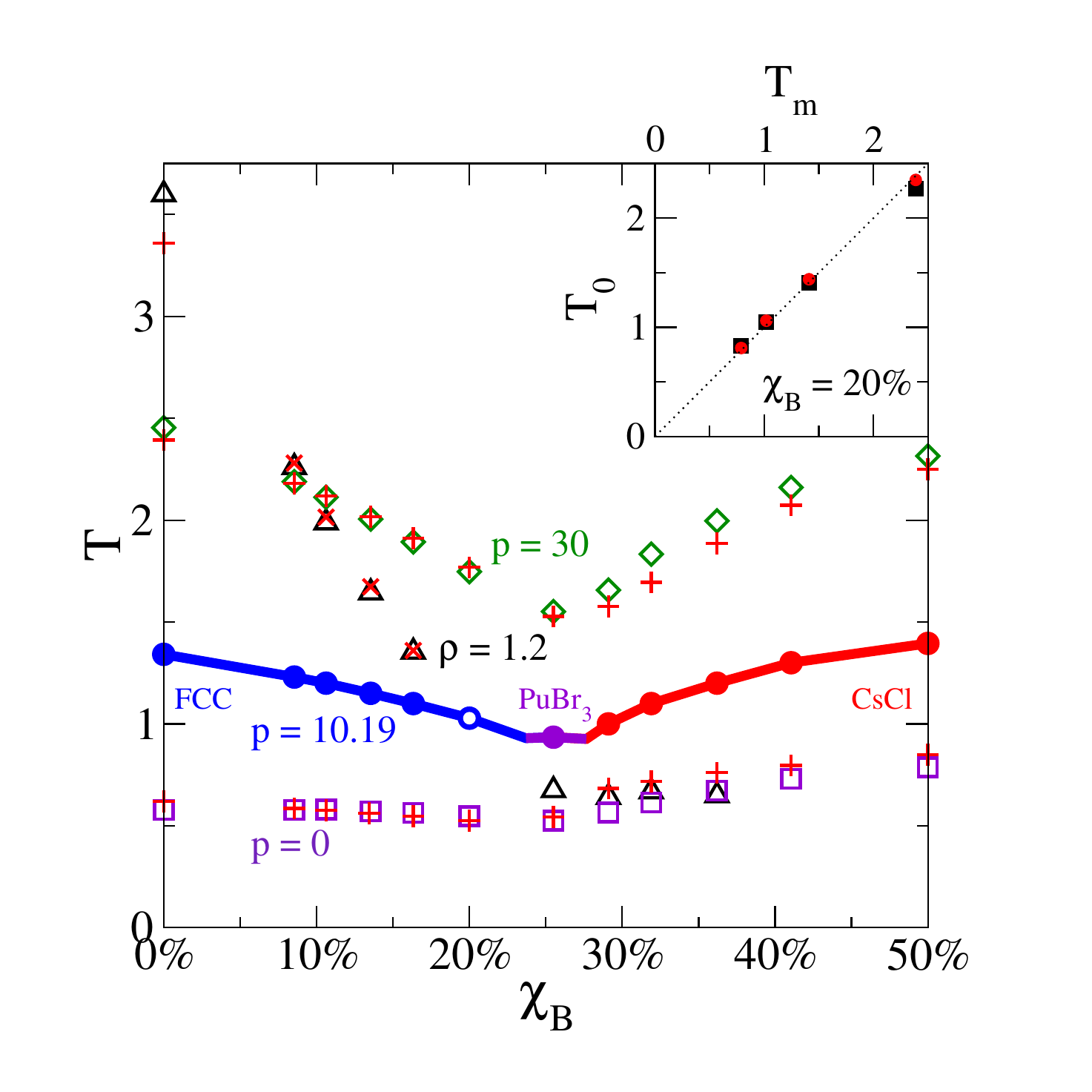} 
		\caption{\label{isoprediction} Isomorph-theory estimated melting temperature as a function of B particle concentration at fixed liquid-phase density $\rho=1.2$ (triangles), as well as at pressures $p=0$ and $p=30$ (squares and diamonds). The circles connected by colored full lines are the results from \fig{pd}, giving the reference states from which the open symbols are predicted. Red crosses are the actual melting temperatures. Inset: Melting temperature $T_m$ versus the onset temperatures $T_0$ identified by Coslovich and Pastore at which the A and B particle dynamics (black and red) change from high-temperature Arrhenius to non-Arrhenius temperature dependence \cite{cos07}. The four points shown represent different densities/pressures of the standard 80/20 KA model.}
	\end{center} 
\end{figure}

\begin{table} 
\begin{tabular}{lr|rcccccc }
   & &$\chi_B^\text{liquid}$ & $T_m$ & $\rho_s$ & $\rho_l$ & $u_s$ & $u_l$ & $\Delta S$ \\
\hline
 FCC &[Fm$\overline{\rm 3}m$; 225]      & 0\%    & 1.341(4) & 1.0535 & 0.9796 & -3.6460 & -2.7971 & 1.177 \\
 PuBr$_3$& [Cmcm; 63] & 25.5\% & 0.935(8) & 1.3515 & 1.2712 & -7.1192 & -6.3170 & 1.367  \\
 Fe$_3$C & [Pnma; 62]& 25.0\% & 0.820(8) & 1.3501 & 1.2950 & -7.1446 & -6.5360 & 1.134 \\ 
 Ni$_3$P & [I$\overline{\rm 4}$; 82]& 24.8\% & 0.767(6) & 1.3582 & 1.2878 & -7.0759 & -6.6409 & 1.102 \\
 Al$_2$Cu& [I$\overline{\rm 4}$/mcm; 140]& 33.4\% & 0.875(8) & 1.4256 & 1.2858 & -7.1720 & -6.9785 & 1.109\\
 CsCl    & [Pm$\overline{\rm 3}$m; 221]& 49.4\%   & 1.394(4) & 1.6427 & 1.4392 & -7.5850 & -5.9561 & 1.798 \\
\end{tabular}
\caption{\label{tableTm} State point data for the liquid in coexistence with different crystalline states at pressure $p=10.19$ at the specified B particle concentrations in the liquid phase (A particle LJ units). Crystal structures are identified by prototype, Hermann-Mauguin space group, and IUCr number. $T_m$ is the melting temperature, $\rho_s$ and $\rho_l$ are the solid and liquid densities, $u_s$ and $u_l$ are the average potential energies per particle of the solid and liquid phases, $\Delta S$ is the constant-pressure entropy of fusion calculated from the parameters and the phase-equilibrium condition $\Delta G=0$.}
\end{table}

The melting line in \fig{pd} is for $p=10.19$, the pressure at which the coexisting liquid for the standard 80/20 KA system has density $1.2$. The melting temperature's variation with B particle concentration at other pressures or at constant liquid-phase density may be estimated from the $p=10.19$ results from the fact that the melting line to a good approximation follows a liquid isomorph \cite{IV,ped16}. The estimate, which makes use of the fact that LJ-type systems are R-simple, i.e., have strong virial potential-energy correlations \cite{ped08,I,dyr14}, works as follows. For any mixture of LJ particles, if the melting temperature at liquid density $\rho_0$ is $T_0$, the melting temperature $T$ at liquid density $\rho$ is approximately given \cite{boh12,ing12a} by

\begin{equation}\label{eq1}
T / T_0
\,=\, (\gamma_0/2 -1)(\rho/\rho_0)^4\,-\,(\gamma_0/2 -2)(\rho/\rho_0)^2\,.
\end{equation}
Here $\gamma_0$ is the density-scaling exponent at $(\rho_0,T_0)$ computed from equilibrium canonical-ensemble fluctuations via $\gamma_0=\langle\Delta U \Delta W\rangle/\langle (\Delta U)^2\rangle$ in which $U$ is the potential energy and $W$ the virial \cite{IV}. At liquid density $1.2$ the predictions based on \eq{eq1} are shown as triangles in \fig{isoprediction}.

If one wishes to estimate from the $p=10.19$ results the melting temperature at other pressures, the following procedure is employed. Along an LJ isomorph the pressure varies as $p/\rho=k_BT + (2w_0-4u_0)(\rho/\rho_0)^4-(w_0-4u_0)(\rho/\rho_0)^2$ in which $w_0$ and $u_0$ are the virial and potential energy per particle at the reference state point $(\rho_0,T_0)$ \cite{V}. \Eq{eq1} is inserted into this expression, resulting in $p(\rho)/\rho = A (\rho/\rho_0)^4-B (\rho/\rho_0)^2 $ in which $A=k_BT_0(\gamma_0/2-1)+2w_0-4u_0$ and $B=k_BT_0(\gamma_0/2-2)+w_0-4u_0$. This relation is inverted numerically and the melting temperature is estimated by inserting $\rho(p)$ into \eq{eq1}. The parameters used in the predictions are given in Table \ref{tabel2}.

The estimated melting temperature at $p=0$ and $p=30$ are shown in \fig{isoprediction} as open squares and diamonds, respectively. These compare reasonably well to the actual melting temperatures (crosses). At the standard 80/20 composition the zero-pressure melting temperature is estimated to be 0.547(8). The actual melting temperature is $0.525(4)$, which is 4\% below the estimate. The largest deviation from the isomorph-theory estimated melting temperatures are found for the CsCl structure. In this case there is a four times larger difference between the liquid and crystal density-scaling exponents than for the A particle FCC crystal, implying that the melting line is less accurately represented as an isomorph \cite{ped16}. For single-component systems a more accurate melting-line theory exists based on a first-order Taylor expansion from the approximate melting isomorph \cite{ped16}.

\begin{table}[b]
	\begin{tabular}{c|ccccc} 
		$\chi_B$ & $T_0$ & $\rho_0$ & $\gamma_0$ & $u_0$ & $w_0$ \\ 
		\hline 
		0.000 &  1.341 &  0.980 &  5.16 &  -4.817 &  9.056  \\ 
		0.085 &  1.230 &  1.062 &  5.15 &  -5.256 &  8.372  \\ 
		0.106 &  1.200 &  1.084 &  5.11 &  -5.372 &  8.204  \\ 
		0.135 &  1.150 &  1.117 &  5.12 &  -5.548 &  7.971  \\ 
		0.163 &  1.100 &  1.151 &  5.10 &  -5.724 &  7.752  \\ 
		0.200 &  1.028 &  1.200 &  5.11 &  -5.974 &  7.444  \\ 
		0.255 &  0.935 &  1.278 &  5.10 &  -6.343 &  7.042  \\ 
		0.291 &  1.000 &  1.305 &  5.13 &  -6.354 &  6.808  \\ 
		0.319 &  1.100 &  1.316 &  5.18 &  -6.263 &  6.641  \\ 
		0.362 &  1.200 &  1.343 &  5.18 &  -6.202 &  6.384  \\ 
		0.410 &  1.300 &  1.375 &  5.19 &  -6.118 &  6.115  \\ 
		0.494 &  1.395 &  1.439 &  5.21 &  -5.953 &  5.691  \\ 
	\end{tabular} 
	\caption{\label{tabel2} Thermodynamic quantities along the $p=10.19$ melting line used for estimating the melting temperatures at constant liquid-phase density, as well as at two other pressures, reported in \fig{isoprediction}.}
\end{table}

\begin{figure} 
	\begin{center} 
		\includegraphics[width=0.5\textwidth]{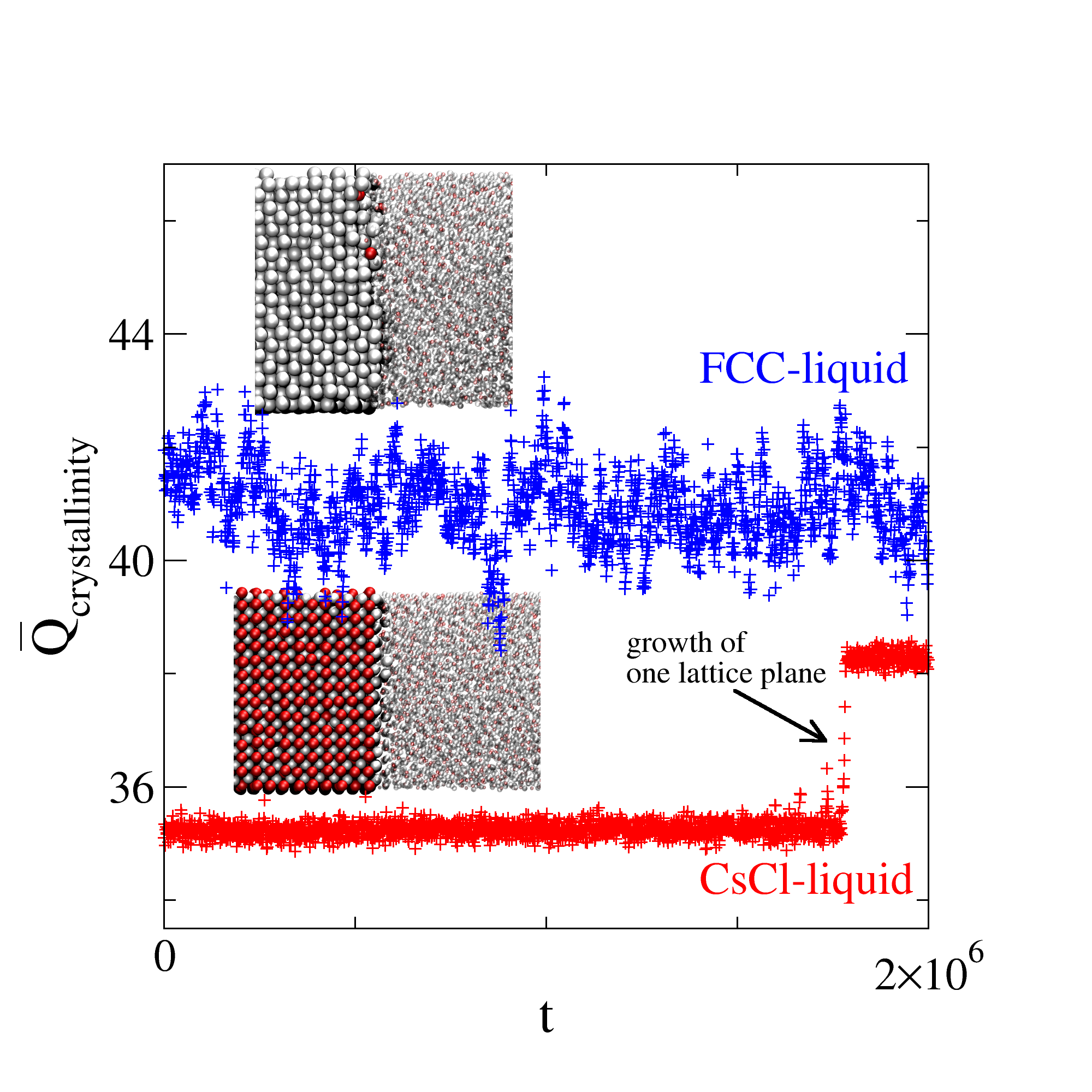} 
		\caption{\label{Q} The crystal order parameter $Q$ \cite{ped13} plotted as a function of time averaged over $10^3$ LJ time units. The blue crosses are from an FCC-liquid coexistence simulation at $T=1.028$, the red crosses are from a CsCl-liquid coexistence simulation at $T=1$. These results indicate that the FCC interface is rough, whereas the CsCl interface is smooth and flat. The latter property impedes growth of the CsCl crystal. Particles with $Q_6>0.25$ are highlighted in the inserted coexistence pictures, where $Q_6$ is the rotational order-parameter \cite{lec08}.} 
	\end{center} 
\end{figure} 

By fitting relaxation times for the standard 80/20 KA system to the frustration-limited-domain theory of the glass transition \cite{tar04}, Coslovich and Pastore determined the  ``onset temperature'' below which the non-Arrhenius temperature dependence of the relaxation time initiates \cite{cos07}. The inset of \fig{isoprediction} compares their data to the melting temperature at different pressures. There is good agreement, which tentatively confirms physical pictures of Sastry and of Yanagishima \textit{et al.} \cite{sas98a,sri,yan17}. Though the identity of onset and melting temperatures may not apply universally, it suggests that a class of liquids exists for which this is the case.

The two main crystal structures have different formation dynamics. \Fig{Q} shows how the crystal order parameter fluctuates at solid-liquid equilibrium. The coexisting FCC-liquid situation is more noisy than the CsCl-liquid case. We interpret the former as reflecting a rough interface, whereas in the latter case the interface is flat, i.e., grows layer by layer \cite{Villain_bog}. This is consistent with general findings \cite{Kin_Proc} since the melting entropy is larger in the latter case (Table \ref{tableTm}).

In summary, we have mapped out the phase diagram of KA-type binary LJ mixtures showing, in particular, that the standard 80/20 KA liquid at density $1.2$ crystallizes at $T=1.028(3)$. Our findings suggest that KA systems more stable against crystallization are those of 26-30\% B particles. This is consistent with recent results of Ingebrigtsen and Royall, who have simulated the density $1.2$ KA viscous liquid with 30\% B particles for months on a GPU cluster without being able to crystallize it \cite{ing_unpub}.

\acknowledgments
This work was supported by the VILLUM Foundation's VKR-023455 and \textit{Matter} (16515) grants.


%\bibliography{jcd}
\end{document}